\documentclass[prl,aps,twocolumn,amssymb,floatfix,floats,showpacs]{revtex4}
\usepackage{epsfig}
\usepackage{color}
\newcommand{\be}{\begin{equation}}
\newcommand{\ee}{\end{equation}}
\newcommand{\bea}{\begin{eqnarray}}
\newcommand{\eea}{\end{eqnarray}}

\def\g{\gamma}

\def\d{\delta}
\def\a{\alpha}
\def\e{\varepsilon}

\def\nn{\nonumber\\}

\def\fr#1{(\ref{#1})}

\def\2t#1#2{\langle\tau_{#1}\tau_{#2}\rangle}

\def\nn{\nonumber\\}

\def\ii{{\rm i}}
\def\d{{\rm d}}
\def\e{{\rm e}}

\begin{document}

\title{Current large deviation function for the open asymmetric
simple exclusion process} 

\author{Jan de Gier$^1$ and Fabian H. L. Essler $^2$}
\affiliation{$^1$ Department of Mathematics and Statistics, The
  University of Melbourne, 3010 VIC, Australia\\ 
$^2$ Rudolf Peierls Centre for Theoretical Physics, University of
Oxford, 1 Keble Road, Oxford, OX1 3NP, United Kingdom} 

\begin{abstract}
We consider the one dimensional asymmetric exclusion process
with particle injection and extraction at two boundaries. The model is
known to exhibit four distinct phases in its stationary state. We
analyze the current statistics at the first site in the low and high
density phases. In the limit of infinite system size, we conjecture an
exact expression for the current large deviation function. 
\end{abstract}


\maketitle
\emph{Introduction.}
One of the main open problems in classical statistical physics is the
formulation and derivation of simple laws that determine macroscopic
quantities in strongly interacting systems far from equilibrium. A
broad class of nonequilibrium systems can be characterized by the
presence of a macroscopic current. An important diagnostic tool of
non-equilibrium behaviour is then provided by the probability
distribution of current fluctuations. The latter is suitably
represented in terms of its moments, which are encoded in the current large
deviation function (LDF). LDFs play an important role in the
application of fluctuation theorems \cite{EvansSearles,GC,JarzCrooks}. 
Microscopic models of interacting particles provide
a useful framework for studying non-equilibrium properties in
current-carrying classical systems and have become a  major subject of
research over the past two decades. One of their main uses is that
their large deviation properties can be derived microscopically, which
furnishes rigorous tests of underlying assumptions in phenomenological
approaches.  \newline
The asymmetric simple exclusion process (ASEP),
describing the asymmetric diffusion of hard-core particles along a 
one-dimensional chain, is one of the best studied paradigms
of non-equilibrium Statistical Mechanics \cite{ASEPreview}. 
The ASEP is of general interest due to its close relation to
growth phenomena \cite{KPZ}, as observed in recent experiments on
electroconvection \cite{Takeuchi}. It is also used as a model of molecular
diffusion in zeolites \cite{HahnKK96}, of biopolymers \cite{biopolymer1}
and sequence alignment \cite{bundschuh}, traffic flow
\cite{ChowdSS00} and quantum dot chains \cite{Oppen}. The exact probability distribution for current
fluctuations for the ASEP on a ring has been known for some time
\cite{LDFring}. In the open boundary ASEP phenomenological
\cite{Ber_etal}, approximate \cite{Oppen} and numerical \cite{Mitsudo} 
treatments have been developed, but the determination of the current LDF
from first principles has been one of the outstanding
problems in the field. Despite considerable effort, the LDF
is only known in the limiting cases of symmetric exclusion
\cite{DerridaDR04} and weak asymmetry \cite{BodD0406}. For the infinite system the time dependence was obtained for total asymmetry in \cite{PraehS}. 

\begin{figure}[t]
\includegraphics[width=0.9\columnwidth]{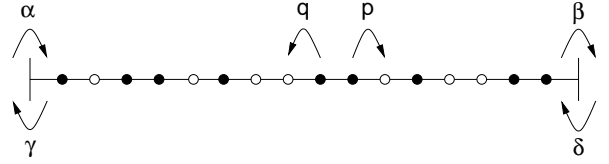}
\caption{Dynamical rules of the ASEP.}
\label{fig:asep}
\end{figure}

\paragraph{Definition of the ASEP.}
At any given time $t$ each site is either occupied by a particle or
empty and the system evolves subject to the following rules. In the
bulk ($i=2,\ldots,L-1$) a particle attempts to 
hop one site to the right with rate $p$ and one site to the left with
rate $q$. The hop is executed unless the neighbouring site is
occupied, in which case nothing happens. On the first and last sites
these rules are modified by allowing particles to enter (leave) 
with rates $\alpha$ ($\gamma$) at site $i=1$ and with rates
$\delta$ ($\beta$) at site $i=L$ respectively, see Figure~\ref{fig:asep}. 

\begin{figure}[t]
\includegraphics[width=0.8\columnwidth]{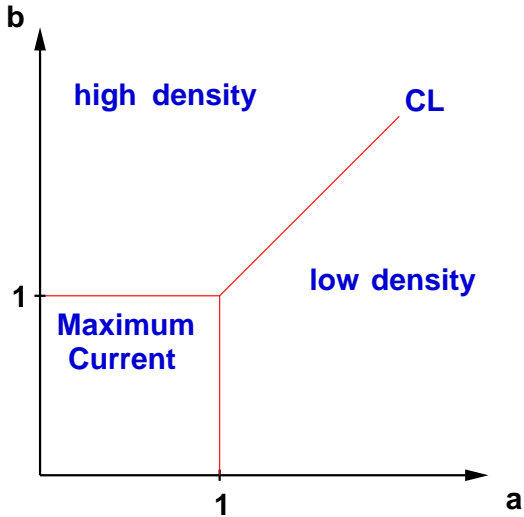}
\caption{Stationary state phase diagram for the ASEP. On the coexistence
  line (CL) a first order phase transition occurs.}
\label{fig:statPD}
\end{figure}

With every site $i$ we
associate a Boolean variable $\tau_i$, indicating whether a particle
is present ($\tau_i=1$) or not ($\tau_i=0$). The state of the system
at time $t$ is then characterized by the probability distribution
$P_t(\tau_1,\ldots,\tau_L)$. The time evolution of $P_t$ occurs
according to the aforementioned rules and is subject to the
master equation    
\be
\frac{\d P_t}{\d t} = M P_t.
\label{eq:Markov}
\ee
Here $M=m_1+m_L+m_{\rm bulk}$ is the ASEP transition matrix whose
eigenvalues have non-positive real parts. The late time behaviour of
the ASEP is dominated by the eigenstates of $M$ with the largest real
parts of the  corresponding eigenvalues \cite{dGE}. The boundary contributions
$m_1$ and $m_L$ describe injection (extraction) of particles 
at sites $1$ and $L$. In the
following we use a more convenient parametrization in terms of the
quantities $a=\kappa^+_{\alpha,\gamma}$, $b=\kappa^+_{\beta,\delta}$,
$c=\kappa^-_{\alpha,\gamma}$, $d=\kappa^-_{\beta,\delta}$, where
\be
\kappa^{\pm}_{\alpha,\gamma} = \frac{
p-q-\alpha+\gamma\pm \left[(p-q-\alpha+\gamma)^2
  +4\alpha\gamma\right]^{\frac12}}{2\alpha}.
\ee
\paragraph{Stationary state properties of the ASEP.}
At late times the ASEP approaches a stationary state. Physical
properties then depend sensitively on the boundary conditions
\cite{PASEPstat}. For $q<p$ one finds four different phases as a
function of the boundary rates as is shown in Fig.\ref{fig:statPD}.

\paragraph{Current Fluctuations.}
We are interested in the probability distribution of the total
time-integrated current $Q_1(t)$, i.e. the net number of particle jumps
between the left boundary reservoir and site 1 in the time interval
$[0, t]$. The moments of the distribution are encoded in the
generating function $\langle \e^{\lambda Q_1(t)}\rangle$, where the
brackets denote an average over all histories.  In this Letter we report an
explicit expression for the quantity 
\be
\label{Elam}
E(\lambda) = \lim_{L\to\infty}\lim_{t\rightarrow\infty} \frac1t \log
\langle \e^{\lambda Q_1(t)}\rangle. 
\ee
This characterizes the asymptotic current distribution which, for an
ergodic system, is not expected to depend on the choice of initial
particle configuration. 

As observed in \cite{LS}, eqn \fr{Elam} implies a large deviation
property for the probability distribution $P(j_1, t)$ of the average
current $j_1 = Q_1(t)/t$ at the first site. The long-time limiting
behaviour is given by $P(j_1, t)\sim \e^{-t \widehat{E}(j_1)}$ where 
$\widehat{E}(j_1) = \max_{\lambda} \left\{ \lambda
  j_1-E(\lambda)\right\}$ is the Legendre transform of $E(\lambda)$.
As a tool to compute the current LDF we introduce a fugacity
$\e^{\lambda}$ conjugate to the current on the first site. The
boundary term $m_1$ then becomes
\be
m_1=\left(\matrix{-\a&\g\e^{-\lambda} \cr \a \e^\lambda&-\g\cr}\right)\otimes \mathbb{I}_{L-1},
\label{h1l}
\ee
and $E(\lambda)$ is equal to the largest eigenvalue of the generalized
``transition matrix'' $M(\lambda)$. 
The spectrum of $M(\lambda)$ obeys a Gallavotti-Cohen symmetry
\cite{EvansSearles,GC,LS,dGE}: the eigenvalues of $M(\lambda')$ and
$M(\lambda)$ are equal when $\lambda'$ and $\lambda$ are related by
$\e^{\lambda'}=abcd q^{L-1}\e^{-\lambda}$.
\paragraph{Summary of Results.}
Our main result is that the generating function \fr{Elam} for current fluctuations at site $1$ in the low and high density phase and for small $\lambda$ and $L\rightarrow\infty$ is of the form
\be
E(\lambda)=(p-q)
\frac{a(\e^{\lambda}-1)}{(1+a)(\e^{\lambda}+a)}.
\label{eq:Elambda1}
\ee
In the high density phase we obtain the same expression with $a$
replaced by $b$. Note that the requirements that $\lambda$ is small
and $L\rightarrow\infty$ explicitly break the Gallavotti-Cohen
symmetry, as this is a duality between small and large negative
$\lambda$. We may use (\ref{eq:Elambda1}) to derive explicit
expressions for the first few cumulants of the local current in terms
of the average bulk density $\rho=1/(1+a)$, 
\bea
&&\lim_{t\to\infty}\frac{\langle Q_1\rangle}{t}=
(p-q)\rho(1-\rho),\nn
&&\lim_{t\to\infty}\frac{\langle Q_1^2\rangle-\langle Q_1\rangle^2}{t}=
(p-q)\rho(1-\rho)(1-2\rho),\nn
&&\lim_{t\to\infty}\frac{
\langle Q_1^3\rangle-3\langle Q_1^2\rangle
\langle Q_1\rangle
+\langle Q_1\rangle^3}{t}=\nn
&&\qquad\qquad(p-q)\Big[\rho-7\rho^2+12\rho^3-6\rho^4\Big].
\eea
The first result reproduces, as expected, the bulk current
\cite{PASEPstat}, while the second moment agrees with the diffusion
constant in the limit $q\to 0$ of completely asymmetric diffusion
\cite{DerridaEM95}.
\paragraph{Derivation.}
In the following we set $p=1$ without loss of generality.
Based on earlier work on the quantum XXZ spin chain
\cite{CaoNepo03}, the generalized ASEP transition matrix
was shown to be diagonalizable using the Bethe ansatz in the case
where the parameters satisfy \cite{dGE,simon}
\be
\left(q^{L/2+k}-\e^{\lambda}\right)\left(\alpha\beta\e^{\lambda} -
q^{L/2-k-1}\gamma\delta\right) =0.
\label{pasepconstr}
\ee
Here $k$ is an arbitrary integer in the interval $|k|\leq L/2$. 
By considering small finite systems we find that the largest
eigenvalue $E(\lambda)$ is described by one of the sets of Bethe
equations given in [\onlinecite{dGE}], which can be cast in the
form
\bea
E &=& \sum_{l=1}^{L/2+k}\frac{\left(1-q\right)^2 z_l}{(1-z_l)(1-qz_l)} \equiv
\sum_{l=1}^{n}\varepsilon(z_l),
\label{eq:pasep_en2}\\
Y_L(z_j) &=& \frac{2\pi}{L} I_j,\quad j=1\,\dots,\frac{L}{2}+k,
\label{eq:logBA}
\eea
where $n=L/2+k$, and $Y_L(z)$ is given by
\bea
\ii\, Y_L(z) &=& g(z)+ \frac1L g_b(z)-\left(1-\frac{n-1}L\right)\ln (-qz) \nonumber\\
&&+\frac1L \sum_{l=1}^{n} K(z_l,z).
\label{eq:pasep_eq2blog}
\eea
Here the functions $g$, $g_{\rm b}$ and $K$ are given by
\bea
g(z)&=&\ln\left[z\frac{(1-qz)^2}{(1-z)^2}\right],
\label{eq:g}\\
g_b(z)&=&
\ln\left[-\frac{1+az}{a+qz}\frac{1+cz}{c+qz}\right]
+\ln\left[-\frac{1+bz}{b+qz}\frac{1+dz}{d+qz}\right]\nonumber\\
&&{}+\ln\left[\frac1z\frac{1-q^2z^2}{1-z^2}\right].
\label{eq:gb}\\
K(w,z) &=&-\ln(w) - \ln \left(\frac{1-qz/w}{1-qw/z} \frac{1-q^2wz}{1-wz}\right).  
\label{eq:K}
\eea
We note that these equations are different from those describing the
low lying excitations of the ASEP \cite{dGE}. 

The constraint \fr{pasepconstr} can be
satisfied for arbitrary $\alpha,\beta,\gamma,\delta$, $q$ and $k$ by
fixing the parameter $\lambda$ characterizing the generating function
to a value among the sequences (S1) $\lambda_n^{(1)}=n\ln(q)$ or (S2) 
$\lambda_n^{(2)}=\ln\left(\gamma\delta q^{n-1}/\alpha\beta\right)$, 
where $n$ is an integer with $0\leq n\leq L$. In order to infer $E(\lambda)$
we employ the following strategy: we set
$\lambda=\lambda^{(j)}_n$ and then determine the ground state energies
$E(\lambda^{(j)}_n)$ of the corresponding generalized transition
matrices. From the sequences of values obtained in this manner we
then conjecture a general expression for $E(\lambda)$.

\emph{Ground State Energy for sequence (S1)}.
Here, the ground state in the low density phase corresponds to a
solution of the Bethe ansatz equations with only $n$ roots
($n=1,2,\ldots$) 
\be
z_j=-\frac{q^{j-1}}{a}+{\cal O}\Big(e^{-\mu_j L}\Big)\ ,\quad j=1,\ldots,n,
\label{zjS1}
\ee
where for large $L$ the $\mu_j$ approach constant values. We have
checked (\ref{zjS1}) against exact diagonalization of small chains 
$(L\leq 14)$ for many values of the boundary rates and $n\leq 5$. We
conjecture that it is correct in general for sufficiently small $n$, 
i.e. $n$ such that $q^{2n}>abcdq^{L-1}=\gamma\delta q^{L-1}/\alpha\beta$.  
The solution \fr{zjS1} is of the form of a maximal \emph{boundary bound
  state}: one root lies exponentially close to a pole of the boundary
phase shift $e^{g_b(z)}$, while pairs of the others lie on poles of
the two-particle phase shift $e^{K(z_k,z_l)}$. The ground state energy
\fr{eq:pasep_en2} becomes 
\be
E=\sum_{j=0}^{n-1} \epsilon\left(-\frac{q^{j-1}}{a}\right) =(1-q)\left(\frac{a}{a+1}-\frac{a}{a+q^{n}}\right).
\ee
Restoring $\lambda$ and $p$ we obtain the result (\ref{eq:Elambda1}). 

%
%

\begin{figure}[ht]
\centerline{\includegraphics[width=0.45\columnwidth]{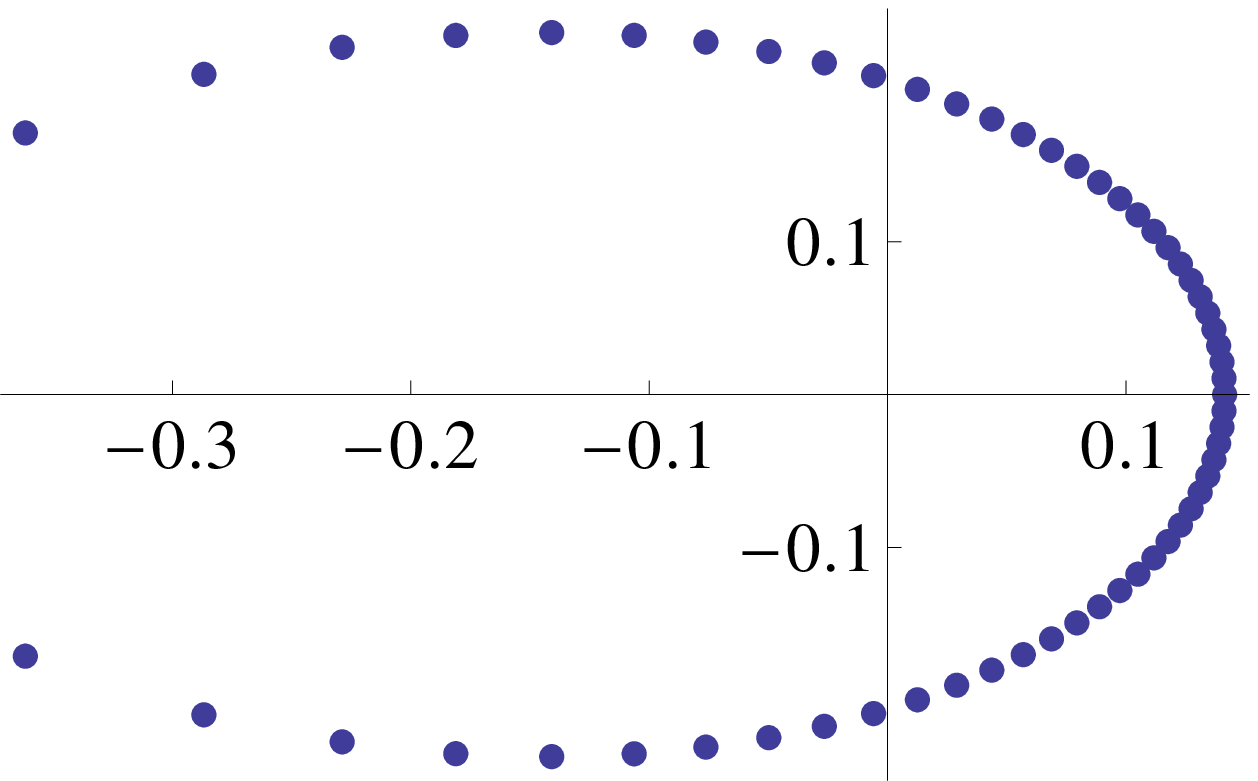}\qquad
\includegraphics[width=0.45\columnwidth]{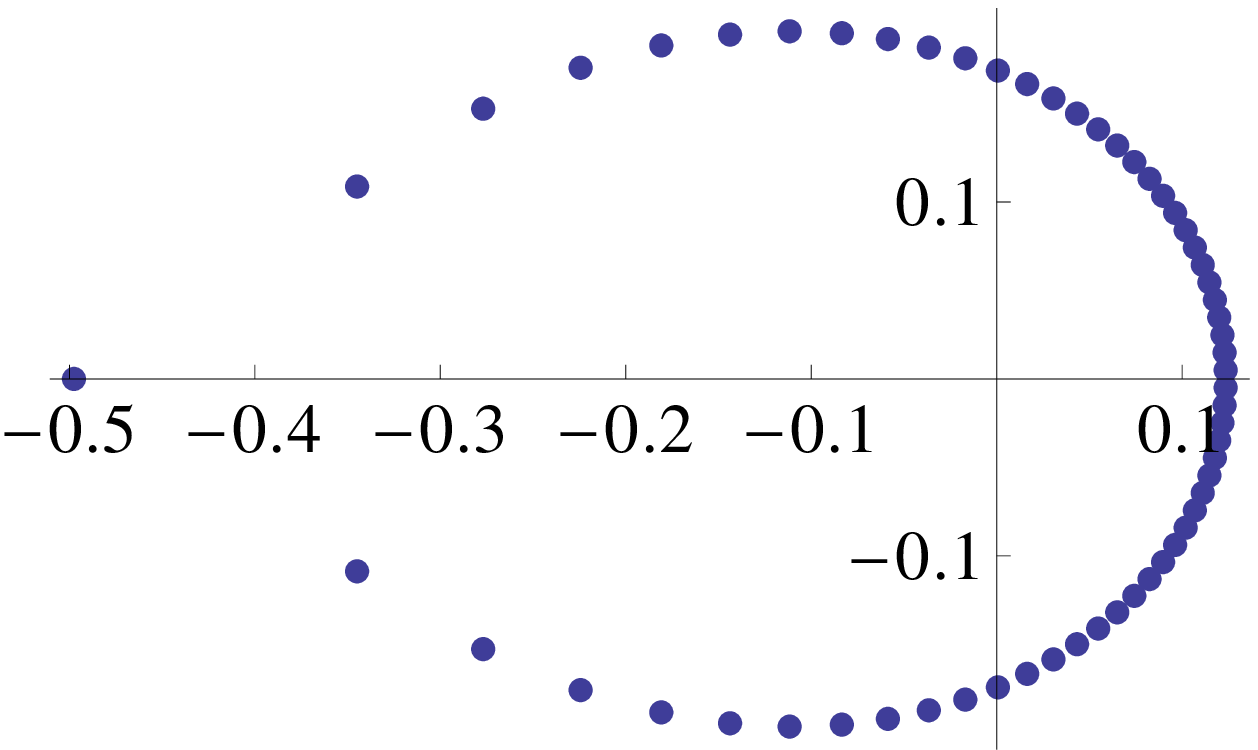}}
\caption{Distribution of reciprocal roots $1/z_j$ for $L=60$. Left: $a=3.45$, $b=1.5$, $c=-0.55$, $d=-0.6$ and $q=0.8$. Right: $a=1.7$, $b=1.6$, $c=-0.55$, $d=-0.6$ and $q=0.9$. Both contours close on the negative real axis as $L$ increases.}.
\label{fig:L=60}
\end{figure}

\emph{Ground State Energy for sequence (S2)}. 
Here the analysis is considerably more involved. The ground state in
the low density phase 
is again given by \fr{eq:pasep_en2}, \fr{eq:logBA}, but now with
$k=L/2-n$. To keep $\lambda^{(2)}_n$ small for $L\gg 1$ we require
$n\ll L$, which corresponds to the number of Bethe roots being $\mathcal{O}(L)$. In the
following we present details for the case $n=1$, other values can be
treated analogously. For $n=1$ there are $L-1$ roots. The ground state
is obtained by choosing 
\be
I_j=-L/2+j,\qquad j=1,\ldots,L-1.
\ee
The corresponding roots lie on a contour that closes as $L\rightarrow
\infty$ on a point $z_{\rm c}$ on the negative real axis, see e.g. the
plot on the left hand side of Figure~\ref{fig:L=60}.  

Following \cite{dGE} we obtain an integro-differential equation for 
the root density $Y_L(z)$ in the limit $L\rightarrow\infty$, valid in
the low and high density phases. Dropping subleading contributions in
$L^{-1}$ we have
\be
\ii\,Y_L(z) = g(z) + \frac{1}{L} g_{\rm b}(z) +\frac{1}{2\pi}
\int_{\xi^-}^{\xi^+} K(w,z) Y'_L(w) \d w.
\label{eq:intY}
\ee
The integral from $\xi^-$ to $\xi^+$ is along the contour formed by the roots, and the end points are fixed by $Y_L(\xi^\pm) = \pm(\pi -\pi/L)$. Equation \fr{eq:intY} may be solved by expanding in powers of $L^{-1}$, i.e.
$Y_L(z) = y_0(z) + y_1(z)/L +\ldots$, $\xi = z_{\rm c} + (\delta + \ii\, \eta)/L +\ldots$,
which upon substitution into (\ref{eq:intY}) yield integro-differential equations for the functions $y_0$ and $y_1$. Once these have been determined the corresponding eigenvalue $E(\lambda^{(2)}_n)$ is obtained from
\be
E =  - \frac{L}{2\pi} \oint_{z_{\rm c}} \varepsilon(z)Y'_L(z)\d z -
\frac{\ii}{\pi} y_0'(z_{\rm c})
\, \eta\, \varepsilon(z_{\rm c})+\dots,
\label{eq:energyMI}
\ee
where we have dropped terms of ${\cal O}(L^{-1})$. Here, the integral is over the closed contour on which the roots lie.

\paragraph{Assumption I: $\lambda>0$, $-1/a$ inside the contour.}
%
This regime corresponds to the case where $e^{\lambda}=abcd>1$, and is defined by assuming that $-1/a$ lies inside the contour of integration and all other poles of $g_{\rm b}$ lie outside. The zeroth order term in the expansion of the counting function can be found as in \cite{dGE}, and is given by 
\be
y_{0}(z) = -\ii \ln \left[ - \frac{z}{z_{\rm c}}
\left(\frac{1-z_{\rm c}}{1-z}\right)^2 \right].
\ee

Under the above assumption, the driving term of the subleading integro-differential equation can be shown to have branch points at $-1/a$ and at $qz_{\rm c}$. The branch point at $-1/a$ results in branch points in $y_1(z)$ at the points $-q^m/a$, $m=0,1,2\ldots$, and likewise for the branch point at $qz_{\rm c}$. As in \cite{dGE}, this suggests a functional form for $y_1(z)$ which may then be obtained explicitly.

Finally, employing the boundary conditions for $\xi^\pm$, it is possible to show that in this regime $\delta=0$, $\eta\, y_0'(z_{\rm c})=\ii\pi$ and that the contour closes at $z_{\rm c}=-bcd$, which agrees well with numerical solutions of \fr{eq:logBA} up to $L=200$. The energy can be computed from \fr{eq:energyMI} and is given by 
\be
E= (1-q) \left( \frac{a}{a+1} - \frac{1}{1-z_{\rm c}}\right),
\label{eq:energy2}
\ee 
which with $z_{\rm c}=-bcd$ is fully consistent with \fr{eq:Elambda1}, and coincides with it when we restore $p$ and $\lambda$ using $\e^\lambda=abcd$. 

\paragraph{Assumption II: $\lambda<0$, $g_{\rm b}$ analytic inside the contour.}
%

A numerical analysis of the case $\e^{\lambda}=abcd<1$ indicates that the roots again lie on a contour, except for isolated roots on the negative real axis. The plot on the right hand side of Figure~\ref{fig:L=60} gives an example with only one such isolated root $z_1\approx-1/a$. Under the assumption that the boundary term $g_{\rm b}$ does not have poles inside the contour, the leading order integro-differential equations may again be obtained explicitly. While the details of the calculation of the energy are slightly different from above, the final result is again \fr{eq:energy2} with $z_{\rm c}=-bcd$, confirming also in this case \fr{eq:Elambda1}.

\paragraph{Conclusions.} In this letter we have presented a
conjecture for the exact current LDF in the high
and low density phases of the ASEP with open boundaries in the limit
of infinite system size. We have presented strong evidence in
support of our conjecture. While the density LDF for the open ASEP has
been derived from microscopic first principles some time ago
\cite{DerridaLS}, the exact determination of its current LDF
has been an important outstanding problem in non-equilibrium
statistical physics. Both quantities are assumed to fully describe the
experimentally accessible macroscopic behaviour of the ASEP \cite{Takeuchi}. 
So far we have not been able to access the coexistence line and the
maximum current phase, where it is necessary to scale the parameter
$\lambda$ in a non-trivial way with system size \cite{LDFring}. In the
maximum current phase we have been able to analyze only the limit
$L\to\infty$ for fixed $\lambda$, in which we obtain the following
result 
\be
E(\lambda)=(p-q)
\tanh (\lambda/4).
\label{eq:Elambda2}
\ee
It would be interesting to see whether further progress can be made
for weak asymmetry, c.f. \cite{WASEPring}.
Finally we note that we have obtained preliminary results
on finite-size corrections to \fr{eq:Elambda1}.

\acknowledgments
We thank K. Mallick and R. Stinchcombe for helpful discussions. This work was
supported by the ARC, the EPSRC under grant EP/D050952/1 and the John
Fell OUP Research Fund.


\begin{thebibliography}{99}

\bibitem{EvansSearles}
D.J. Evans, E.G.D. Cohen and G.P. Moriss,
Phys. Rev. Lett. {\bf 71}, 2401 (1993); 
D.J. Evans and D.J. Searles, Phys. Rev. E {\bf 50}, 1645 (1994). 
%
\bibitem{GC} G. Gallavotti and E.G.D. Cohen, Phys. Rev. Lett. {\bf
  74}, 2694 (1995); J. Stat. Phys. {\bf 80}, 931 (1995).
%
\bibitem{JarzCrooks} C. Jarzynski, Phys. Rev. Lett. {\bf 78}, 2690
  (1997); 
G. E. Crooks, J. Stat. Phys. {\bf 90}, 1481 (1998);
J. Kurchan, J. Phys. {\bf A31}, 3719 (1998).
%
\bibitem{ASEPreview} 
B.~Derrida, Phys. Rep. {\bf 301}, 65 (1998);
 G.M.~Sch\"utz, {\it Phase Transitions and Critical Phenomena} {\bf
   19} (Academic Press, London, 2000).  
%
\bibitem{KPZ} M. Kardar, G. Parisi and Y.C. Zhang,
  Phys. Rev. Lett. {\bf 56}, 889 (1986);
K. Johansson,   Comm. Math. Phys. {\bf 209}, 437 (2000); 
C.A. Tracy and H. Widom, Comm. Math. Phys. {\bf 290}, 129 (2009);
J. Math. Phys. {\bf 50}, 095204 (2009);
T. Sasamoto and H. Spohn, Phys. Rev. Lett. {\bf 104}, 230602 (2010); 
P. Calabrese, P. Le Doussal and A. Rosso, Eur. Phys. Lett. {\bf 90},
20002 (2010);
V. Dotsenko, Eur. Phys. Lett. {\bf 90}, 20003 (2010).
%
\bibitem{Takeuchi} 
K.A. Takeuchi and M. Sano, Phys. Rev. Lett. {\bf 104}, 230601 (2010).
%
\bibitem{HahnKK96} 
K. Hahn, J. K\"arger, and V. Kukla,
  Phys. Rev. Lett. {\bf 76}, 2762 (1996).
\bibitem{biopolymer1} 
G.M. Sch\"utz, Europhys. Lett. {\bf 48}, 623 (1999).
%
\bibitem{bundschuh} 
R. Bundschuh, Phys. Rev. E {\bf 65}, 031911 (2002).
%
\bibitem{ChowdSS00} 
D. Chowdhury, L. Santen and A. Schadschneider,
  Phys. Rep. {\bf 329}, 199 (2000).
%
\bibitem{Oppen}
T. Karzig and F. von Oppen, Phys. Rev. B {\bf 81}, 045317 (2010).

%
\bibitem{LDFring}
B. Derrida and J.L. Lebowitz, Phys. Rev. Lett. {\bf 80}, 209 (1998);
D.S. Lee and D. Kim, Phys. Rev. E {\bf 59}, 6476 (1999);
C. Appert-Rolland {\sl et. al.}, 
Phys. Rev. E \textbf{78}, 021122 (2008);
S. Prolhac, J. Phys. A {\bf 41}, 365003 (2008); J. Phys. A {\bf 43},
105002 (2010).
%
\bibitem{Ber_etal} L. Bertini {\sl et. al.},
Phys. Rev. Lett. {\bf 87},  040601 (2001); 
Phys. Rev. Lett. {\bf 94}, 030601 (2005).
%
\bibitem{Mitsudo} 
T. Mitsudo and S. Takesue, arXiv:1012.1387.
%
%
\bibitem{DerridaDR04} 
B. Derrida, B. Dou\c cot and P.E. Roche, J. Stat. Phys. {\bf 115}, 717 (2004).

\bibitem{BodD0406} 
T. Bodineau and B. Derrida, J. Stat. Phys. {\bf 123}, 277 (2006).

\bibitem{PraehS} M. Praehofer and H. Spohn, \textit{In and out of equilibrium}, ed. V. Sidoravicius, Progress in Probability \textbf{51}, 185 (Birkhauser Boston, 2002). 

\bibitem{dGE}
J. de Gier and F.H.L. Essler, Phys. Rev. Lett. \textbf{95}, 240601
(2005); J. Stat. Mech. P12011 (2006); J. Phys. A \textbf{41}, 485002
(2008).
%
\bibitem{PASEPstat}
B. Derrida, M. Evans, V. Hakim and V. Pasquier, J. Phys. A {\bf 26},
1493 (1993);
G. Sch\"utz and E. Domany, J. Stat. Phys. {\bf 72}, 277 (1993);
S. Sandow, Phys. Rev. E {\bf 50}, 2660 (1994);
F.H.L. Essler and V. Rittenberg, J. Phys. A {\bf 29}, 3375 (1996);
T. Sasamoto, J. Phys. A {\bf 32}, 7109 (1999), J. Phys. Soc. Jpn {\bf
  69}, 1055 (2000);
R.A. Blythe {\sl et. al.}
, J. Phys. {\bf A33}, 2313 (2000);
M. Depken and R. Stinchcombe, Phys. Rev. Lett. {\bf 93}, 040602
(2004).

\bibitem{LS} 
J.L. Lebowitz and H. Spohn, J. Stat. Phys. {\bf 95}, 333 (1999).
%
\bibitem{DerridaEM95} 
B. Derrida, M.R. Evans and K. Mallick, J. Stat. Phys. {\bf 79}, 833 (1995).
%
\bibitem{CaoNepo03}
 R.~I.~Nepomechie, J. Stat. Phys. {\bf 111}, 1363 (2003);
 J. Phys. A {\bf 37}, 433 (2004);
R.~I. Nepomechie and F.~Ravanini, J. Phys. A {\bf 36}, 11391 (2003);
J.~Cao {\sl et. al.}, 
Nucl. Phys. B {\bf 663}, 487 (2003);
H. Frahm {\sl et. al.}, J. Phys. {\bf A44}, 015001 (2011).
%
\bibitem{simon}
D. Simon, J. Stat. Mech. P07017 (2009).
%

\bibitem{DerridaLS} B. Derrida, J.L. Lebowitz and E.R. Speer, Phys. Rev. Lett. {\bf 89}, 030601 (2002); J. Stat. Phys. {\bf 110}, 775 (2003).

\bibitem{WASEPring} S. Prolhac and K. Mallick, J. Phys. A {\bf 42},
  175001 (2009). D. Simon, arXiv:1011.3590.
%

\end{thebibliography}
\end{document}